\documentclass[twocolumn,trackchanges,floatfix]{aastex631}
\shorttitle{UV-bright stars in NGC\,2420}
\graphicspath{{./}{figures/}}

\begin{document}
\title{UOCS. XIII. Study of the FUV bright stars in the open cluster NGC 2420 using ASTROSAT}
\author{R.K.S. Yadav}  
\affiliation{ Aryabhatta Research Institute of Observational Sciences, Manora Peak, Nainital 263002, India.} 

\author{Arvind K. Dattatrey}
\affiliation{ Aryabhatta Research Institute of Observational Sciences, Manora Peak, Nainital 263002, India.}
\affiliation{ Deen Dayal Upadhyay Gorakhpur University, Gorakhpur, Uttar Pradesh 273009,India.} 

\author{Geeta Rangwal}
\affiliation{South-Western Institute for Astronomy Research, Yunnan University, Kunming 650500, People’s Republic of China}
\author{Annapurni Subramaniam}
\affiliation{Indian Institute of Astrophysics, Koramangala, Bangalore 560034, India.}

%

\author{D. Bisht}
\affiliation{Indian Centre for Space Physics, 466 Barakhola, Singabari road, Netai Nagar, Kolkata 700099, India.}

\author{Ram Sagar}
\affiliation{Indian Institute of Astrophysics, Koramangala, Bangalore 560034, India.}

\

\email{arvind@aries.res.in}
\email{rkant@aries.res.in}


\begin{abstract}
 We present the study of four FUV bright stars in the field of open cluster NGC 2420 using the Ultra Violet Imaging Telescope (UVIT) mounted on AstroSat. The three stars 525, 527, and 560 are members, while star 646 is a non-member of the cluster. To characterize and determine the parameters of these stars, multi-wavelength spectral energy distributions (SEDs) are analyzed using UV, optical, and IR data sets. For all four FUV bright stars, a two-component SED model fits well. Our findings indicate that two stars, 525 and 560, are binary BSS systems. These binary BSS systems may have formed in a tertiary system due to mass transfer from an evolved outer tertiary companion. Star 527 is a binary system of a BSS and an extremely low-mass (ELM) white dwarf, while star 646 is a binary system of a horizontal branch star and an ELM white dwarf. The effective temperatures, radii, luminosities, and masses of the two ELMs are (10250, 11500) K, (0.42, 0.12) R$_\odot$, (1.61, 0.23) L$_\odot$, and (0.186, 0.170) M$_\odot$, respectively. The star 527 could be a post-mass transfer system and may have originated through the Case A/B mass-transfer process in a low-density environment. The cooling age of the ELMs is $<$ 1 Myr, indicating that they have only recently formed. 
\end{abstract}



\section{Introduction}
\label{sec:intro}
Open cluster ultraviolet (UV) imaging is frequently dominated by hot, bright UV-bright stars. Wide-field UV imaging is crucial for gathering a complete sample of a cluster's hot population. The UV suppresses the dominating cool-star population and highlights the hot stars, ensuring that all hot populations in the cluster are detected. Blue straggler stars (BSSs) predominate among the hot population in open clusters.

BSSs are observed to be brighter and bluer than the main sequence turn-off lying along an extrapolation of the main sequence in the color-magnitude diagrams (CMDs) of  star clusters \citep{1953AJ.....58...61S}. The possible mechanisms of BSS formation are mass transfer in a binary system \citep{1964MNRAS.128..147M} and direct stellar collisions \citep{1976ApL....17...87H}. In recent studies, it was also proposed that BSS formed from the merger of main sequence stars previously in a hierarchical triple system as a result of the eccentric Kozai-Lidov mechanism \citep{2009ApJ...697.1048P,2014ApJ...793..137N}, which has a significant role in BSS formation in open clusters \citep{2023A&A...672A..81L}. Many studies have been presented to describe the BSS populations and their formation scenarios in open and globular clusters \citep{2009Natur.462.1028F, 1986AJ.....92.1364M,2023ApJ...943..130D} 

{NGC 2420 ($\alpha_{J2000}$ = $07^{h}38^{m}23^{s}.0$, $\delta_{J2000}$ = 
 $21^{\circ}34^{\prime}24^{\prime\prime}$) is a metal-poor open cluster with [Fe/H]$ = -0.26$ situated at a distance of 3.08 kpc in the northern Galactic hemisphere. It is located 19$^o$ above the galactic plane in the third quadrant. Its age is 2.0 Gyr \citep{2008A&A...481..149D} and interstellar extinction $E(B-V)$ is 0.03 mag. It contains a number of BSSs and red-giant branch (RGB) stars. The above parameters are taken from WEBDA\footnote{https://webda.physics.muni.cz/navigation.html}.

Due to its richness and age, NGC 2420 has been the subject of several studies. The richness allows us to find stars in many rarer stages of stellar evolution using cluster CMDs.  \cite{2008A&A...481..149D}   performed the first UV study using the GALEX data. Their study aimed to search for the hot-evolved objects using GALEX and SDSS data. \cite{2019AJ....158...35S} observed this cluster using UVOT and compared the UV CMD with the model. \cite{2000AJ....120.1384V}   observed the cluster using a WFPC2 camera mounted on HST and identified eight white dwarf candidates. Additionally, UV photometry can be utilized to explore the extended main-sequence turn-off (eMSTO) of the open clusters. The cause of eMSTO is still a mystery, but it is believed to originated from varying rotation rates among stars. In the present investigation, we aim to characterize the Far-UV (FUV) bright sources detected by UVIT in the direction of the open cluster NGC 2420. 
The FUV band of UVIT is highly sensitive in detecting hot star populations. Combining FUV to Near-IR bands enables us to detect hot objects in binary systems where the un-evolved cooler companion dominates the optical light.     

 \cite{Thompson_2021} estimated a binary fraction of 41\% in the cluster NGC 2420. A binary sequence is presented by \citet{2001MmSAI..72..707P} in this cluster. According to \cite{2016MmSAI..87..505G}, the binary properties (frequency, orbital elements and companion masses, and evolutionary states) of the blue stragglers are the most crucial diagnostic for determining their origins. Their findings demonstrate the close relationship between blue stragglers and binaries. They discovered compelling evidence that many blue stragglers in the clusters NGC 188, NGC 2682 (M67), and NGC 6819 formed via mass transfer from an evolved star to a main-sequence star. However, all viable formation channels likely produce the blue straggler populations simultaneously.

\section{Data Sets and their reduction}
\label{sec:data}
The UV data of the open cluster NGC 2420 were downloaded from the UVIT archive. The observations were done on 30$^{th}$ April 2018 in one FUV (F148W) filter of UVIT under the proposal ID A04$_{-}$075 T05 (PI: Prof. Ram Sagar). 
The observed data set is reduced using a customized software package CCDLAB \citep{2017PASP..129k5002P}. 
The corrections for satellite drift, flat-field distortion, fixed pattern noise, and cosmic rays are applied by the CCDLAB. The telescope, instruments, and preliminary calibration are described in detail in \cite{2016SPIE.9905E..1FS}  and \cite{2017AJ....154..128T}. The DAOPHOT routine provided by \cite{1987PASP...99..191S} performs the point spread function (PSF) photometry. The photometric error as a function of the F148 magnitude is shown in Fig. \ref{fig:FUV_error}. The photometric error is $\le$ 0.1 mag at 20 mag in the F148 band. The photometric depth in the F148 band is $\sim$ 23.5 mag with 0.8 mag error.

Archival UV data collected with the Ultraviolet Optical Telescope (UVOT) in uvw1, uvm2 and uvw2 filters are also used in this analysis. The raw data from the HEASARC archive\footnote{https://heasarc.gsfc.nasa.gov} was processed using the HEA-Soft\footnote{https://heasarc.gsfc.nasa.gov/docs/software/heasoft} pipeline. The telescope's specifications and photometric calibration of UVOT data can be found in \cite{2008MNRAS.383..627P}. The corrections for exposure maps and auxiliary spacecraft data, as well as the science images that were geometrically corrected for sky coordinates, were applied. The method described by \cite{2014AJ....148..131S} was applied to reduce the UVOT/Swift data. On the other hand, optical photometric  data are taken from the GAIA DR3 catalogue for the CMDs.


\begin{figure}
\centering
\includegraphics[width=6cm, height=4cm]{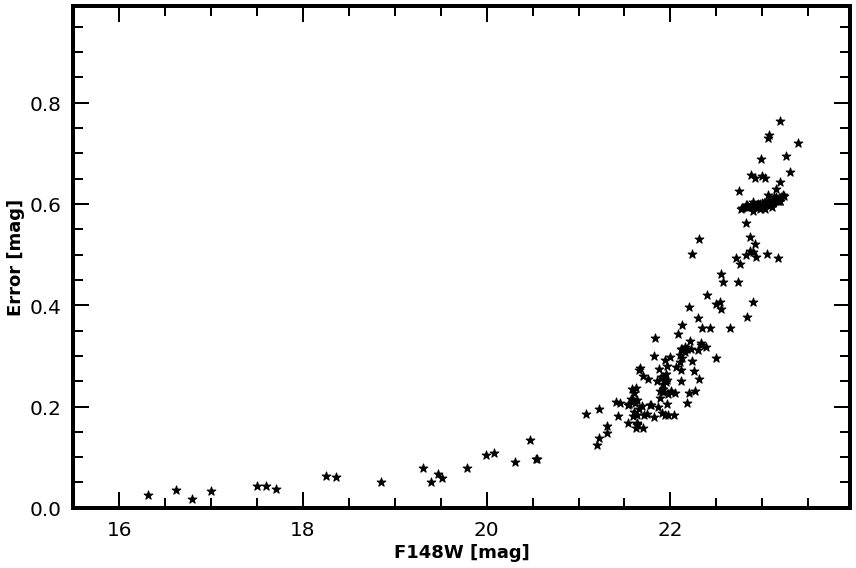}
\caption{The photometric errors in F148 UVIT filter. }
\label{fig:FUV_error}
\end{figure}

\begin{figure}
    \centering
    \includegraphics[width=6cm, height=4cm]{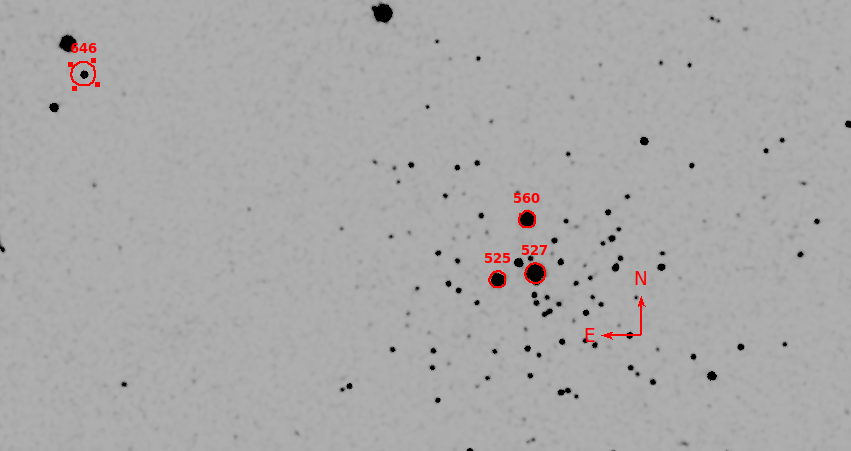}
    \caption{The identification chart of the four UV bright stars shown with the star numbers in FUV image. The field-of-view is 11$^{\prime}\times11^{\prime}$.}
    \label{fig:chart}
\end{figure}

\section{UV and Optical Colour Magnitude Diagrams} \label{sec:CMD}
\subsection{The selection of FUV-bright stars}

To identify the bright stars in the FUV range, we constructed the cluster's FUV-optical, optical, and FUV-NUV CMDs. To find the membership probability of the stars, we used the proper motion data taken from the GAIA DR3 catalogue. The process adopted to estimate the membership probability is described by \cite{2013MNRAS.430.3350Y}. The method is based on the detailed comparison between the cluster-like stars' kinematic distribution and the field's. We constructed the CMDs using the stars having a membership probability of more than 90\%, and the results are shown in Fig. \ref{fig:CMD}. In the UV plane, hot stars are located above the turn-off point, which distinguishes them from other stars in the cluster. The eMSTO is visible in the UV CMDs, possibly due to differences in star rotation rates \citep{2015MNRAS.450.3750M}. An isochrone of age 2 Gyr with a red line is overplotted in all the CMDs with the cluster's parameters listed in Sec. \ref{sec:intro}. The ZAMS shown with a blue line is taken from Padova\footnote{http://stev.oapd.inaf.it/cgi-bin/cmd}. The isochrone fits well in each CMD. The tip of the MS is detected at an FUV (F148) magnitude of $\sim$21.5 in the FUV-optical CMD. A similar CMD is also presented in \cite{2021JApA...42...89J}. In the top left CMD, four FUV bright stars, identified by their numbers 527, 525, 560, and 646, are clearly visible above the turn-off point. They are depicted with a diamond shape. These stars are also positioned above the cluster's turn-off point in other CMDs presented in Fig. \ref{fig:CMD}. 

The present analysis shows a membership probability greater than 90\% for 527, 525, 560, and 646. In the GALAH spectroscopic survey \citep{2021MNRAS.503.3279S}, 527, 525, and 560 have more than 90\%, while the star 646 has 17\% membership probability. The cluster radius is 5$\arcmin$, derived by \citet{2006AJ....132.1669S} 
 using wide field (50$\arcmin$$\times$50$\arcmin$) optical photometry. The stars 527, 525, and 560 are located within the cluster radius as their distance from the center is less than 5$^{\arcmin}$. The position of star 646 is 10$\arcmin$.9 from the cluster center. Hence, this star is out of the cluster region. Based on the membership probability provided in the GALAH survey and position in the image, we assume that stars 527, 525, and 560 are members and that star 646 is a non-member of the cluster.

\begin{figure*}
\centering
\includegraphics[width=0.4\textwidth]{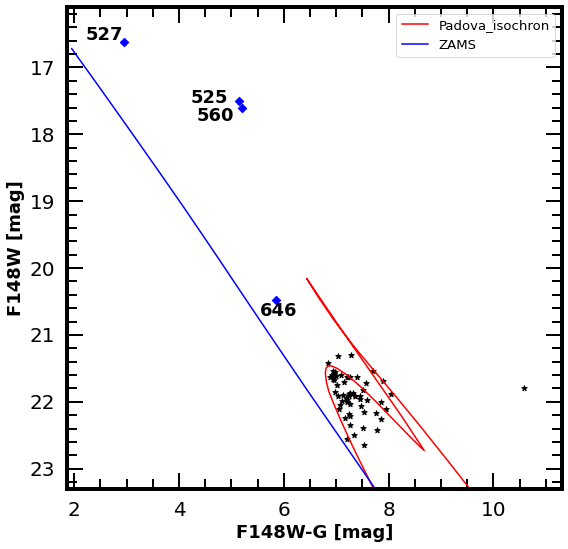} 
\includegraphics[width=0.4\textwidth]{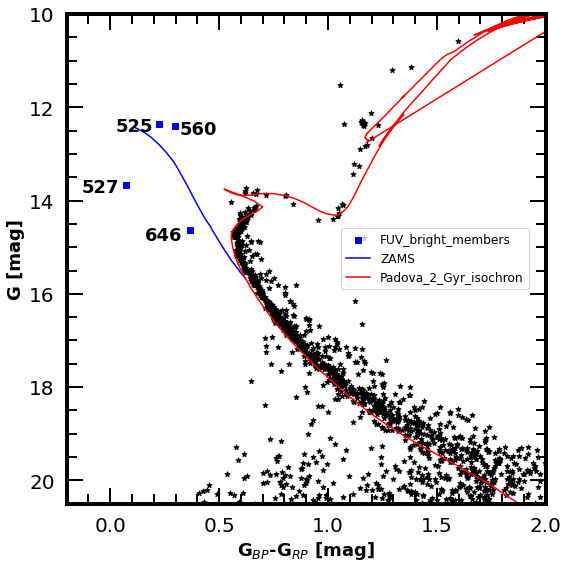} 
\centering
\includegraphics[width=0.4\textwidth]{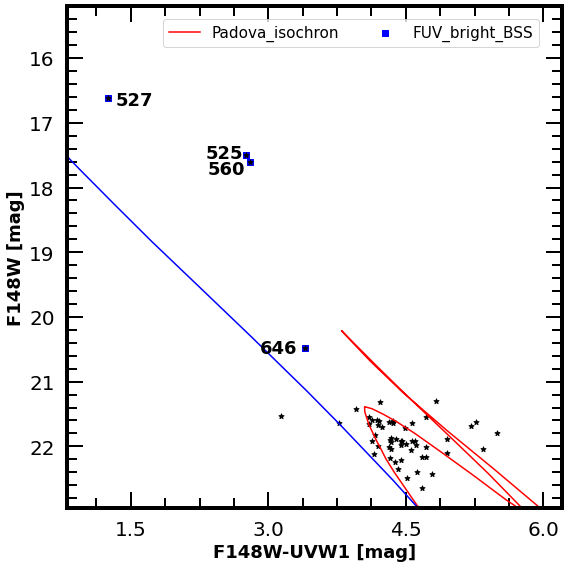}
\includegraphics[width=0.4\textwidth]{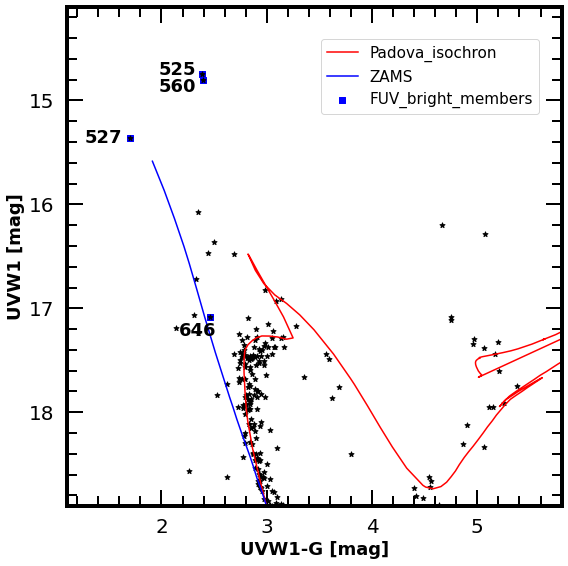}

\caption{The top left, and top right panels show the F148W vs (F148W $-$ G) and G vs ($G_{BP}-G_{RP}$) CMDs, whereas bottom left and bottom right panels show the F148W vs (F148W-UVW1) and UVW1 vs (F148W-UVW1), respectively. All the FUV bright stars are shown with their identification numbers. The Padova isochrone of age 2 Gyr is overplotted with a red line. The ZAMS taken from Padova is shown with a blue line.}
\label{fig:CMD}
\end{figure*}

\section{Spectral Energy Distributions}
\begin{figure}
\setkeys{Gin}{width=\linewidth,height= 4cm}
\includegraphics[scale=1.0]{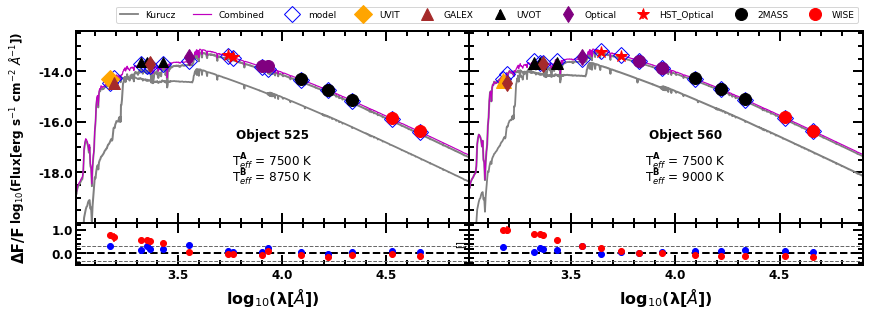}
\includegraphics[scale=1.0]{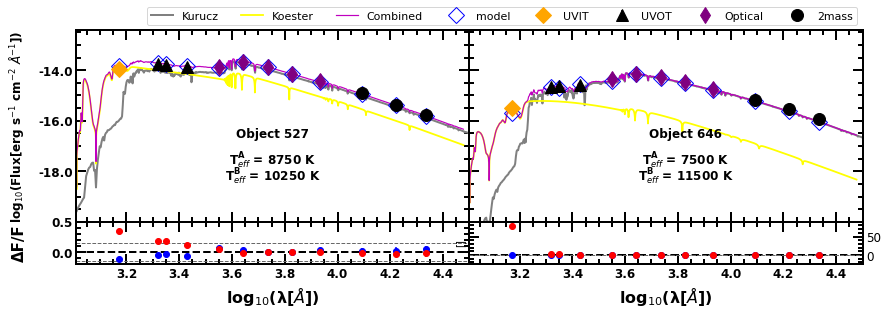}
\hfill

\caption{The SEDs of four FUV bright stars (525, 560, 527, and 646). The T$_{eff}$ of the cool (A) and hot (B) components are displayed in each SED. The UVIT, UVOT, GALEX, optical (Stetson catalogue), HST, 2MASS, and WISE data  points are represented with diffrent colors shown in legend, while the model points are represented by open blue rhombus. The two Kurucz models, Koester model and combined spectra are represented by grey, yellow and magenta, respectively.  $\Delta F/F$ is the fractional residuals.
}
\label{fig:SED}
\end{figure}

We constructed spectral energy distribution (SED) diagrams to examine the characteristics and nature of 527, 525, 560, and 646 stars. The FUV UVIT data are combined with UVOT, GALEX, the optical data from Stetson\footnote{https://www.canfar.net/storage/list/STETSON/Standards} and HST, 2MASS, and WISE data to cover a wide 
range of wavelengths in the SEDs. The virtual observatory tool VOSA (VO SED Analyzer, \citep{2008A&A...492..277B}) has been used for the SED and their analysis.  To compute the synthetic photometry for a selected theoretical model, VOSA uses filter transmission curves. The synthetic photometry is then compared to the observed data, and the best fit is determined using the $\chi^2$ minimization test.  The $\chi^2_r$ is defined as:
\begin{equation}
    \chi^2_r = \frac{1}{N-n_p}\sum_{i=1}^{N}\left(\frac{(Y_{i,o}-M_dY_{i,m})^2}{\sigma_{i,o}^2}\right)
\end{equation}
   Where, Y$_{i,o}$ is the observed flux,   \textbf{ $\sigma_{i,o}$ } is the observational error in the flux, Y$_{i,m}$ is the theoretical flux predicted by the model, and M$_{d}$ is the multiplicative dilution factor calculated as M$_{d}$ = $(R/D)^2$, where $R$ and $D$ are the radius of the object and the distance from the observer, respectively. In addition to $\chi^2_r$, VOSA also provides two visual goodness of fit parameters, Vgf and Vgf$_{b}$. It is recommended that Vgf be less than 20-25 and Vgf$_{b}$ is less than 10-15 for a good fitting. 

In VOSA, two-component flux fitting is also possible using a linear combination of two theoretical models. The measured flux, designated by $F_{obs}(x)$, looks like this:
\begin{equation}
    F_{obs}(x) = M_{d1}\cdot F_{m1}(x) + M_{d2}\cdot F_{m2}(x)
\end{equation}
Theoretical fluxes from components 1 and  2 are denoted by $F_{m1}(x)$ and $F_{m2}(x)$, while $M_{d1}$ and $M_{d2}$ are multiplicative dilution factors.

The SED of the four FUV bright stars is shown in Fig. \ref{fig:SED}. The observed data points with error bars are represented by different colours indicated in the figure. The model data points are denoted by open blue rhombus.
First, we fit the  \cite{2003IAUS..210P.A20C}  model with different temperatures and surface gravity but with fixed metallicity [Fe/H]= 0 dex. The fundamental parameters of the cluster listed in Sec. \ref{sec:intro} are assumed for the stars 527, 525, and 560 in the fitting. The distance of star 646 is determined using the parallax (0.466 mas) taken from the GAIA EDR3 catalogue. The effective temperature and log g  are taken in the range of 5000 $-$ 50000 K and  3$-$5 dex, respectively. The SED fitting is also shown in Fig. \ref{fig:SED}. For the single-component SED, we employed the Kurucz atmospheric model, as depicted with grey curves. The residuals between the fitted model and the measured fluxes, normalized by the observed flux, are shown with red points in the bottom panel of each SED. The dashed horizontal lines at $\pm$0.3 (30\%) represent the threshold for residuals. The residual diagrams demonstrate that the residual flux is within $\pm$ 0.3 in optical and IR regions. This shows that the Kuruz model fits well in these wavelength regions. On the other hand, residual flux is greater than 0.3 for the data points in the UV region. This indicates that all four FUV bright stars have UV excess, and the SEDs may be best fit by a combination of hotter and cooler components.

\begin{table*}
\centering
 \caption{The best-fit parameters of the cool and hot components. Here, T$_{eff}$ is the effective temperature in K, $\chi^2_r $ is reduced $\chi^2$, luminosity, radius, and mass are in the solar unit, V$_{gf}$ \& V$_{gfb}$ are the visual goodness of fit, N$_{fit}$ is the total number of points taken into account during the fitting. The mass and age of cool and hot companions are listed in the last two columns.}
 \begin{tabular}{ccccccccccccc}
 \hline
    Name & RA       & DEC   & T$_{eff}$     &Luminosity & Radius &	Log g &	$\chi^2_r$ &	V$_{gf}$ &	V$_{gfb}$ & N$_{fit}$ &Log(age) &Mass \\ 
&(deg)&(deg)&(K)&(L$_{\odot}$)&(R$_{\odot})$&&&&&&(yr)&(M$_{\odot}$) \\
    \hline

525A  & 114.6223& 21.5752 &7500$^{+500}_{-250}$  	   &   69.43$^{+7.92}_{-0.07}$ & 4.65$^{+0.30}_{-0.10}$ &3.0		&11.13 &  6.51  &	1.85 &17/18    &8.9& 2.18 \\  
525B  &&                                  &9000$^{+500}_{-500}$       &   16.68$^{+0.76}_{-3.01}$ & 1.68$^{+0.35}_{-0.15}$ &5.0		&11.13 &  6.51  &   1.85   & 17/18&$\sim 8.8$ & $\sim$ 1.76      \\
\\
527A  &114.609	&21.5778 & 8750$^{+250}_{-250}$    & 26.26$^{+2.50}_{-0.61}$    &  2.21$^{+0.17}_{-0.01}$  & 4.0  & 5.28      & 10.2& 1.04 & 15/16    &8.9& 1.93  \\ 
527B  &               &               &10250$^{+250}_{-750}$      & 1.61$^{+0.17}_{-0.20}$  & 0.42$^{+0.02}_{-0.01}$ &  9.5  &  5.28 & 10.2    & 1.04  & 15/16     &$<$6.0&0.186 \\ 
\\


560A  &114.612	&21.5961&7500$^{+250}_{-250}$  	   &   70.18$^{+5.16}_{-0.51}$ & 4.46$^{+0.40}_{-0.10}$ &4.0		&24.71&  6.07  &	1.95 &17/18  &8.9&2.18   \\ 
560B  & &                             &9000$^{+500}_{-750}$        &   12.50$^{+0.10}_{-1.24}$ & 1.45$^{+0.06}_{-0.80}$ &4.0		&24.71&  6.07  &   1.95   &17/18 &$\sim 8.6$&$\sim 1.71$   \\               
\\

646A  &114.7737	&21.645& 7500$^{+125}_{-125}$    &  4.36$^{+0.95}_{-0.16}$	&	1.24$^{+0.03}_{-0.10}$	&3.5	&8.61&	2.11&0.24      &    15/16	    & ..  &..   \\
646B  &           &     &11500$^{+250}_{-1000}$   &  0.23$^{+0.07}_{-0.20}$	&   0.12$^{+0.26}_{-0.01}$	           &9.0   &8.61 &	2.11&0.24  	   &    15/16       & $<$6.0  &0.16-0.17  \\ 
\hline
\end{tabular}
\label{tab:parameters}
\end{table*}

To account for the UV excess in these stars, we fitted two$-$component SEDs as shown in Fig. \ref{fig:SED}.
We again used the Kurucz stellar atmospheric model \citep{1997A&A...318..841C,2003IAUS..210P.A20C} of higher temperature to fit the hotter components of 525 and 560 and Koester model for 527 and 646. The Koester WD \citep{2010MmSAI..81..921K} model is represented by yellow curves. The range of T$_{eff}$ and log g of the Koester model is 5000 - 80000 K and 6.5 - 9.5 dex, respectively. The combined spectra are shown with magenta curves. The procedure used to fit the combined spectra is described above. The fractional residuals for the combined SEDs are shown as blue points in the bottom panels. For the combined SEDs, the residuals are within 0.3 on each wavelength. The estimated fitting parameters $\chi^2_r$, V$_{gf}$, and V$_{gfb}$ listed in Table \ref{tab:parameters} and the fractional residuals for the combined spectra, indicate that the spectrum fits well for all wavelengths.

Fitting the two components yields the fundamental parameters of the cool and hot components of the FUV bright stars. The estimated parameters are shown in Table \ref{tab:parameters}.  The uncertainties in Table 1 are estimated using a statistical approach described in VOSA. The estimation is carried out by running a 100 iteration Monte Carlo simulation. The star IDs with "A" and "B" represent the cool and hot components, respectively.

\subsection{Age and mass of the components}
The age and mass of each component of  525, 527, and 560 are derived by comparing their positions to theoretical isochrones retrieved from  Padova isochrone \citep{2008A&A...482..883M} using the H-R diagram depicted in Fig. \ref{fig:HR}. The SED-estimated surface temperature and luminosity of each component are considered to construct the H-R diagram. The components A and B are represented by blue and red symbols. The isochrones from 0.8 to 2 Gyr are plotted with grey lines by using the cluster input parameters provided in Sec. \ref{sec:intro}. Based on the comparison of isochrones, we can determine that component A of 525, 527, and 560 have an age of $\sim$ 0.8 Gyr and masses of $\sim$ 2.18, 1.91, and 2.18 M$_{\odot}$, respectively.  The age of BSSs can be determined by the time of mass transfer from a companion star during their formation. The presently derived age indicates the time since the BSS formation. The ages and masses of the cool component are listed in Table \ref{tab:parameters}.  

The hot components B of 525, 527, 560 and 646 are plotted in Fig. \ref{fig:HR}. The hot component of 525 and 560 falls on the Padova isochrones. The age and mass of these components are derived as $\sim$ 0.6 Gyr and 2.76 and  1.76 M$_{\odot}$, respectively. The hot component of 527 and 646 falls below the MS.  To determine this component's age and mass,  we plotted the ELM WD model of different masses, taken from \cite{2016A&A...595A..35I} and represented with magenta curves. The cooling age and mass of 527B and 646B are $<$1 Myr and $\sim$ 0.186 and 0.16 M$_{\odot}$, respectively. The mass and cooling age of the hot components is reported in Table \ref{tab:parameters}.

\section{Results and discussion}\label{sec:results}

We explored four FUV bright stars in the direction of an open cluster NGC 2420 using the UVIT, UVOT, HST, 2MASS, and WISE data sets. Three stars (525,527 and 560) are identified as cluster members, while one star (646) is not a member of the cluster. We constructed the SEDs utilizing VOSA to determine their physical properties and establish whether or not they had any possible companions. The fitting of single component SED to observed data points reveals the UV excess in all four stars. To accommodate UV excess, we fitted the SED with two components. The combined spectra of cool and hot components fit nicely to all four stars. Our analysis shows that stars appear to be part of binary systems. Detailed descriptions of each star are provided below.

{\bf Star 525 and 560}: The positions of 525 and 560 in the optical and FUV CMDs (\ref{fig:CMD}) indicate that these are BSSs. Upon analysis using SED, it has been determined that both stars are binary systems. The cool and hot components of 525 have a luminosity of 69.43 and 16.68 L$_{\odot}$, respectively. Their temperatures are 7500 K and 9000 K, while their radii are 4.64 and 1.68 R$_{\odot}$. On the other hand, the cool and hot components of 560 have a luminosity of 70.18 and 12.50 L$_{\odot}$, respectively. Their temperatures are 7500 K and 9000 K, while their radii are 4.46 and 1.45 R$_{\odot}$. Based on the characteristic parameters, we can deduce that the hot components of these stars may be BSS. Therefore, on the basis of SED analysis, we can infer that 525 and 560 are binary BSS systems. The T$_{eff}$ and log g values of the star 525 are given as 8144 K and 3.9 in the LAMOST\footnote{https://dr7.lamost.org/} 
survey. These values are very close to the present derived values for the cool component.


{\bf Star 527}: From examining the CMDs (\ref{fig:CMD}), we can conclude that this star is a BSS. By analyzing the SED, it has been determined that the star is binary system. The cool component has a luminosity of 26.26L$_{\odot}$, a temperature of 8750 K, and a radius of 2.21 R$_{\odot}$ while the hot component has a luminosity of 1.61 L$_{\odot}$, a temperature of 10250 K, and a radius of 0.42 R$_{\odot}$. Considering these parameters, we can deduce that the cool component is a BSS, whereas the hot component may be an ELM WD.


{\bf Star 646}: This star has been identified as a horizontal branch (HB) star in the LAMOST catalogue. The T$_{eff}$ and log g are given as 7571 K and 3.9 in the catalogue. The SED analysis shows that this star is a binary system. The SED fitting provides the luminosity, temperature, and radius values for the cool component as 4.36 L$_{\odot}$, 7500 K, and 1.24 R$_{\odot}$, respectively. The hot component's values are 0.23 L$_{\odot}$, 11500 K, and 0.12 R$_{\odot}$. The estimated parameters for the cool component are similar to those of an HB star identified in the LAMOST catalogue. Based on the parameters of the hot component, there is a strong indication that this may be an ELM.

\subsection{Formation pathway of the sources}

Based on our findings, we believe that objects 525 and 560 are the binary BSS systems. Table \ref{tab:parameters} indicates that both components of the star 525 and 560 have an age of $\sim$ 1 Gyr. The mass ratio of the components of 525 and 560 is $\sim$ 1.2. Both the components of 525 and 560 have different T$_{eff}$. The mass of these systems is more than the turn-off mass ($\sim$ 1.2 M$_\odot$) of the cluster. The similar ages of the components of 525 and 560 indicate that these binary formed in the same period.  These findings suggest that 525 and 560 are twin BSS binary systems with nearly the same mass but differing T$_{eff}$. 

The twin BSS systems 525 and 560 investigated in this study are similar to the binary BSS system 7782 discovered by \cite{2015ASSL..413...29M} in the old open cluster NGC 188 using spectra. \cite{2019ApJ...876L..33P} suggested a mechanism for creating the BSS binary system 7782.  Based on the simulation, they presented a formation scenario for twin equal-mass BSS. The proposed scenario involves mass transfer from an evolved outer tertiary companion; part of this mass is accreted by the inner binary via a circumbinary disk, while the rest escapes through the second and third Lagrangian points in the potential of the triple system.   

The masses of the components of 525 and 560 are comparable.  According to \cite{2019ApJ...876L..33P}, the BSS in the inner binary produced in tertiary systems should have similar masses regardless of their initial masses.  This is because the lowest mass star often accretes fastest, as its orbital velocity and distance relative to the circumbinary disk is typically the smallest.  They also anticipate that twin BSS generated in tertiary systems will be more common in younger clusters with ages $\le$ 4-6 Gyr.  This is because clusters with an MS turn-off mass $\le$ 1.2 M$_\odot$ have a convective envelope \citep{1991ApJS...76...55I,1995/mnras/275.3.828}, and a radiative envelope for the doner in a mass-transferring binary ensures stable accretion onto the accretor. Therefore, this type of BSS binary system can be found in NGC 2420 (age $\sim$ 2 Gyr).  

Based on our findings, we infer that the system 527 is a binary system. The cool component is BSS, while the hot component is ELM WD.  ELMs are commonly found in binary form in the universe \citep{2019ApJ...886...13J,2019ApJ...883...51R,2020JApA...41...45S,2022MNRAS.511.2274V}. These binary systems can be formed through case A/B mass transfer in binary systems \citep{1991ApJS...76...55I,1995/mnras/275.3.828}, Therefore, mass transfer is required for the creation of ELM WDs in tight binary systems where the companion tears away the ELM WD progenitor’s envelope and the low-mass core fails to ignite the He core. The companion star gained mass and converted into a BSS during the mass loss. 

Our analysis reveals that star 646 is a binary system. The cool component is an HB (particularly RHB star) star, while the hot companion is an ELM WD. After comparing with the evolutionary tracks of ELM WD, it was found that the ELM has a mass of $\sim$0.170 M$_\odot$ and is expected to evolve into a helium-core WD. An ELM white dwarf candidate is expected to form via mass transfer in a binary system. The age of the ELMs shows that they have recently formed. This type of system has been found by \citet{2021ApJ...923..162R} in the galactic globular cluster NGC 2298.

\begin{figure}
\centering
\includegraphics[scale=0.3]{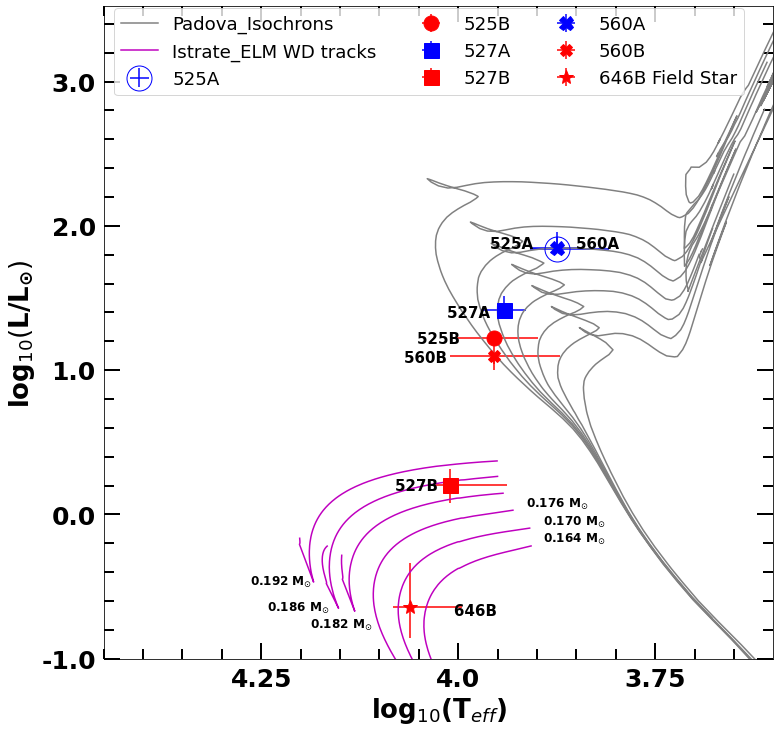}
\caption{The HR diagram of the cool and hot components. The cool and hot components are represented by red and blue points, respectively. The gray and magenta curves represent the Padova isochrones, and Istrate ELM tracks  for different masses, respectively. }
\label{fig:HR}
\end{figure}

\section{Summary and Conclusions}\label{sec:concludions}

In the direction of the open cluster NGC 2420, four FUV bright stars have been identified. The three stars, 525, 527, and 560, are identified as members, and one star, 646, is a non-member of the cluster. The findings are derived from UVIT AstroSat observations, and archival data from UVOT, HST, GAIA EDR3, and the WISE catalogue. We can draw the following conclusions from this study.

\begin{enumerate}

\item To characterize the FUV bright stars, we plotted the SEDs using VOSA. It is evident from the SED analysis that all four sources displayed an excess of FUV flux. To account for the FUV excess, we fitted a two-component SED model.

\item  The components of 525 and 560 fit well with the Kuruz model. The parameters obtained from the two-component SED model can be found in Table \ref{tab:parameters}. Based on our SED analysis, it is conclusive that stars 525 and 560 are binary BSS systems. The other two stars, 527 and 646 have cool and hot companions. The Kuruz model is used for cool companions, while the Koester model is used for hot companions. The cool companion of the star 527 is a BSS, while the hot companion may be ELM WD. On the other hand, the cool companion of 646 is an HB star, whereas the hot companion may be an ELM WD.

\item The twin BSS systems 525 and 560 may have formed through mass transfer from an evolved tertiary star in a triple system. The other two systems, 527 and 646, may have originated due to the mass transfer procedure in binary systems. The cooling age of the hot companions of the 527 and 646 systems implies that they were created very recently.  

\end{enumerate}


\section{ACKNOWLEDGEMENTS}
I would like to express my thanks to Vikrant Jadhav for their help throughout the entire process. 
This study used Topcat (Taylor 2005), Aladin sky atlas (Bonnarel et al. 2000; Boch \& Fernique 2014), Matplotlib (Hunter 2007), and NumPy (Van Der Walt et al. 2011).
I am also thankful to the reviewer for their thoughtful comments and suggestions, which improved the quality of the manuscript. \\
This publication uses data from the Astrosat mission of the Indian Space Research Organisation (ISRO), archived at the Indian Space Science Data Centre (ISSDC). This research made use of VOSA, developed under the Spanish Virtual Observatory project supported by the Spanish MINECO through grant AyA2017-84089. This research also made use of TOPCAT (Taylor 2005, 2011), Aladdin(),  matplotlib (Hunter 2007), and NUMPY (van der Walt, Colbert \& Varoquaux 2011). This research also made use of NASA´s Astrophysics Data System (ADS NASA).

\bibliography{sample631}{}
\bibliographystyle{aasjournal}

\end{document}